\pdfoutput=1
\RequirePackage{ifpdf}
\ifpdf % We~are running pdfTeX in pdf mode
\documentclass[pdftex]{sigma}
\else
\documentclass{sigma}
\fi

\numberwithin{equation}{section}

\newtheorem*{Theorem*}{Theorem}
\newtheorem*{Conjecture*}{Conjecture}

 { \theoremstyle{definition}

 }

\usepackage{dcolumn}% Align table columns on decimal point
\usepackage{bm}% bold math
\usepackage{xcolor}
\usepackage{eepic}

\begin{document}

\allowdisplaybreaks

\newcommand{\arXivNumber}{2112.01735}

\renewcommand{\PaperNumber}{039}

\FirstPageHeading

\ShortArticleName{Doubly Exotic $N$th-Order Superintegrable Classical Systems Separating}

\ArticleName{Doubly Exotic $\boldsymbol{N}$th-Order Superintegrable Classical\\ Systems Separating in Cartesian Coordinates}

\Author{\.{I}smet YURDU\c{S}EN~$^{\rm a}$,
Adri\'an Mauricio ESCOBAR-RUIZ~$^{\rm b}$\newline
and Irlanda PALMA Y MEZA MONTOYA~$^{\rm b}$}

\AuthorNameForHeading{\.{I}.~Yurdu\c{s}en, A.M.~Escobar-Ruiz and I.~Palma y~Meza Montoya}

\Address{$^{\rm a)}$~Department of Mathematics, Hacettepe University, 06800 Beytepe, Ankara, Turkey}
\EmailD{\href{mailto:yurdusen@hacettepe.edu.tr}{yurdusen@hacettepe.edu.tr}}

\Address{$^{\rm b)}$~Departamento de F\'isica, Universidad Aut\'onoma Metropolitana-Iztapalapa,\\
\hphantom{$^{\rm b)}$}~San Rafael Atlixco 186, M\'exico, CDMX, 09340 M\'exico}
\EmailD{\href{admau@xanum.uam.mx}{admau@xanum.uam.mx}, \href{cbi2153013099@izt.uam.mx}{cbi2153013099@izt.uam.mx}}

\ArticleDates{Received December 18, 2021, in final form May 16, 2022; Published online May 27, 2022}

\Abstract{Superintegrable classical Hamiltonian systems in two-dimensional Euclidean space $E_2$ are explored. The study is restricted to Hamiltonians allowing separation of variables $V(x,y)=V_1(x)+V_2(y)$ in Cartesian coordinates. In particular, the Hamiltonian $\mathcal H$ admits a polynomial integral of order $N>2$. Only doubly exotic potentials are considered. These are potentials where none of their separated parts obey any linear ordinary differential equation. An improved procedure to calculate these higher-order superintegrable systems is described in detail. The two basic building blocks of the formalism are non-linear compatibility conditions and the algebra of the integrals of motion. The case $N=5$, where doubly exotic confining potentials appear for the first time, is completely solved to illustrate the present approach. The general case $N>2$ and a formulation of inverse problem in superintegrability are briefly discussed as well.}


\Classification{70H06; 70H33; 70H50}

\section{Introduction}%\label{intro}
For a classical Hamiltonian system with $n$ degrees of freedom, the existence of $n$ integrals of motion in involution is required to make it integrable in the
Liouville sense. These integrals must be well-defined functions in the phase space and functionally independent. On the other hand, a superintegrable system possesses $k$ additional integrals of motion being $k=n-1$ the maximum possible number. The concept of superintegrability can be defined both in classical and quantum mechanics and it has been studied extensively for a very long period of time. The outcome of such a long
period of research activity has far reaching consequences both in mathematical and physical points of view. There exist several exhaustive review
articles in literature which describe the history and current status of this topic \cite{Millerebook, MillerPostWinternitz:2013}.

Starting with a spherically symmetric standard Hamiltonian (i.e., the potential $V=V(r)$ being velocity and spin independent), there exist only
two superintegrable systems, namely the Kepler--Coulomb and the harmonic oscillator. Actually, these two potentials are exactly the ones
which appear in the celebrated Bertrand's theorem \cite{Bertrand1873,Goldstein}. Superintegrability of the Kepler--Coulomb problem is due to the
existence of the conserved Laplace--Runge--Lenz vector \cite{Goldstein, Laplace, Lenz, Runge} whilst in the case of the harmonic oscillator is a consequence of the existence of the quadrupole Jauch--Fradkin tensor \cite{Fradkin, Jauch}.

The systematic investigation of superintegrability has been initiated by Pavel Winternitz and his collaborators in 1965 \cite{Fris}. They
first considered quadratic superintegrability in Euclidean spaces and the subject has been subsequently developed into many directions by several authors since then. For example, its close relation with multiseparability was studied in detail in the references \cite{Evans.b, Evans.a, Fris, Kalnins.a, Makarov, Miller.a},
the search for superintegrable systems in $2$- and $3$-dimensional spaces of constant and nonconstant curvature has been carried out in
the works \cite{Grosche1, Grosche2, Grosche3, Kalnins.a, Kalnins.d, Kalnins.c,Kalnins.b, Kalnins.e,Kalnins.f, Miller.b} and their generalizations to $n$-dimensions
have been analyzed in the papers \cite{Kalnins.i, Kalnins.h, Rodriguez}.

Another important research direction in this field is to consider the Hamiltonians with magnetic field and/or spin. Superintegrability with
magnetic field was first explored in the articles \cite{Berube, Dorizzi} and much recently developed in the articles \cite{BKSnobl, MarSnoblW1,
MarSnoblW2}. The systematic investigation of integrability and superintegrability for systems involving particles with spin
was initiated in the reference \cite{wy1} and subsequently all the rotationally invariant superintegrable systems
in {$E_3$} were classified in the articles \cite{wy3, wy2, YTW}.
On the other hand, spin dependent superintegrable systems were studied in the works \cite{Nikitin1, Nikitin2} for matrix potentials simulating charged or neutral fermions with non-trivial dipole moment in the presence of an electric field.

Still another interesting direction is to go beyond quadratic superintegrability, i.e., the gene\-ral theory of higher-order superintegrability. Initial pioneering works were the articles of
Drach \cite{Drach1,Drach2}, where $10$ potentials allowing third-order integrals of motion were announced. However, much later it was shown that $7$
of these potentials are actually reducible, the third-order integral is the Poisson commutator of two second-order integrals \cite{Ranada, Tsiganov:2000}.
Once again, the systematic investigation of higher-order superintegrability, in particular the third-order one has been initiated by Pavel Winternitz and his collaborators in
the articles \cite{g, gW, Marquette.b, Marquette.a, Popper, TPW2010}.
Almost around the same time higher-order symmetry operators were calculated for
the Schr\"{o}dinger ope\-ra\-tor and the determining equations for the corresponding integrals of motion appeared in~\cite{Nikitin3}. Never\-the\-less, it was soon realized that the analysis became very complicated and some new
ways of approaching to the problem of higher-order superintegrability have to be considered.

After the publication of the seminal paper ``{An infinite family of solvable
and integrable quantum systems on a plane}'' \cite{TTWquantum},
the direction of the research has been thoroughly shifted to {higher-order} integrability/superintegrability
\cite{PostWinternitz:2010, PostWinternitz:2015, TTWclassical}. Moreover, in order to make them more easily tractable, new techniques and methods have been implemented in the study of higher-order {integrable} and superintegrable systems \cite{Chanu1, Chanu2}.

From our point of view one of the main issues on higher-order superintegrability is the classification of the superintegrable potentials. In the case of 2D separable potentials in Cartesian coordinates, an $N$th-order superintegrable system appeared for the first time in~\cite{Thompson}, where the existence of nonlinear equations for the potential which makes the general problem much more complicated was stressed as well. Recently in 2018, by means of a systematic study an infinite 2-parametric family of superintegrable potentials embracing those found in~\cite{GungorKNN:2014, Thompson} was presented in the paper \cite{Grigoriev} by Grigoriev and Tsiganov. Their key element to construct polynomial integrals of motion is the
addition theorems for the action-angle variables, especially the Chebyshev theorem applied to integrals on differential binomials (see also~\cite{Tsiganov:2008B,Tsiganov:2008}). Such an elegant approach has the advantage that it uses the action-angle variables which play a~fundamental role in classical mechanics.

However, unlike the present direct approach some systems can be missed and the gene\-ra\-li\-zation to the quantum case is not straightforward. The explicit list of all $N$th-order 2D (polynomial) superintegrable potential separating in Cartesian coordinates is far to be complete.

Throughout many recent research activity in
the field of higher order superintegrability,
it has been clarified that three types of potentials can occur,
namely the standard, the doubly exotic and singly exotic potentials. Standard ones are solutions of a linear compatibility condition for the determining equations that govern the existence of a higher-order polynomial integral of motion. For doubly exotic potentials this linear compatibility condition is satisfied trivially, it~is identically zero, and the potentials satisfy non-linear equations. These classes of potentials,
appearing in classical and quantum superintegrable systems
have been studied both in Cartesian and polar coordinates
\cite{AW, AMJVPW2015,AMJVPWIY2018, EWY, MSW}.

The aim of this work is to establish in detail general properties of $N$th-order superintegrable classical systems
that allow separation of variables in Cartesian coordinates. It can be considered as the classical counterpart of the general study on quantum superintegrable systems treated in~\cite{AMGen}. However, unlike the latter, in this work we also study the algebra of the integrals of motion and provide exhaustive results for the case $N=5$. In particular, it contains the classical analogues of all the quantum doubly exotic potentials obtained in~\cite{AW} explicitly. We~emphasize that for doubly exotic potentials, unlike the doubly standard ones, the limit from the quantum to the corresponding classical system (i.e., $\hbar \rightarrow 0$) is singular for all the cases studied in the present work. Thus, the corresponding quantum and classical solutions are not connected at~all. The Painlev\'e property characterizing the relevant determining equations in the quantum systems is completely lost in the classical case. In addition, a formulation of inverse problem in superintegrability is briefly discussed as well.

In the present article we focus on 2D classical Hamiltonian systems that are separable in Cartesian variables $(x,y)$
and they also admit an extra polynomial integral of order $N>2$. The generic Hamiltonian is given by
\begin{gather}
{\mathcal H} = {\mathcal H}_1(x) + {\mathcal H}_2(y) \equiv \frac{1}{2}\big(p_1^2 + p_2^2\big) + V_1(x) + V_2(y),
\label{Hcart}
\end{gather}
where $p_i$, $i=1,2$, are the canonical momenta conjugate to $x$ and $y$, respectively. It describes a~two-dimensional particle with unit mass $m=1$ moving in the potential $V(x,y)=V_1(x)+V_2(y)$. Thus, the phase space is four-dimensional. These systems are trivially second-order integrable because in addition to the Hamiltonian (\ref{Hcart}) they admit, for any $V_1(x)$ and $V_2(y)$, another 2nd-order symmetry of the form
\begin{gather}
{\mathcal X} = {\mathcal H}_1(x) - {\mathcal H}_2(y) = \frac{1}{2}\big(p_1^2 - p_2^2\big) + V_1(x) - V_2(y),
\label{X}
\end{gather}
which Poisson commutes (i.e., $\{{\mathcal H},{\mathcal X}\}_{\rm PB}=0$) with the Hamiltonian (\ref{Hcart}). The existence of an $N$th-order third integral $\mathcal Y$, makes the system $N$th-order superintegrable (more integrals of motion than degrees of freedom). In this case, the system is maximally superintegrable. Notice that ${\mathcal H}$ (\ref{Hcart}) is ${\mathcal{S}}_2$-invariant under the permutation $x \Leftrightarrow y $ whilst the integral $\mathcal X$ is anti-invariant.

From a physical point of view we are looking for 2D potentials $V(x,y)=V_1(x)+V_2(y)$ for which all the bounded trajectories are closed and periodic. It is worth mentioning that (\ref{Hcart}) can also be interpreted as the Hamiltonian of the relative motion of a two-body problem on the plane with translational invariance. In this case, $(x,y)$ are nothing but the Cartesian coordinates of the relative vector ${\bf r}={\bf r}_1-{\bf r}_2\equiv (x,y)$ between the two bodies.

\begin{figure}[h]\centering
\includegraphics[scale=0.3]{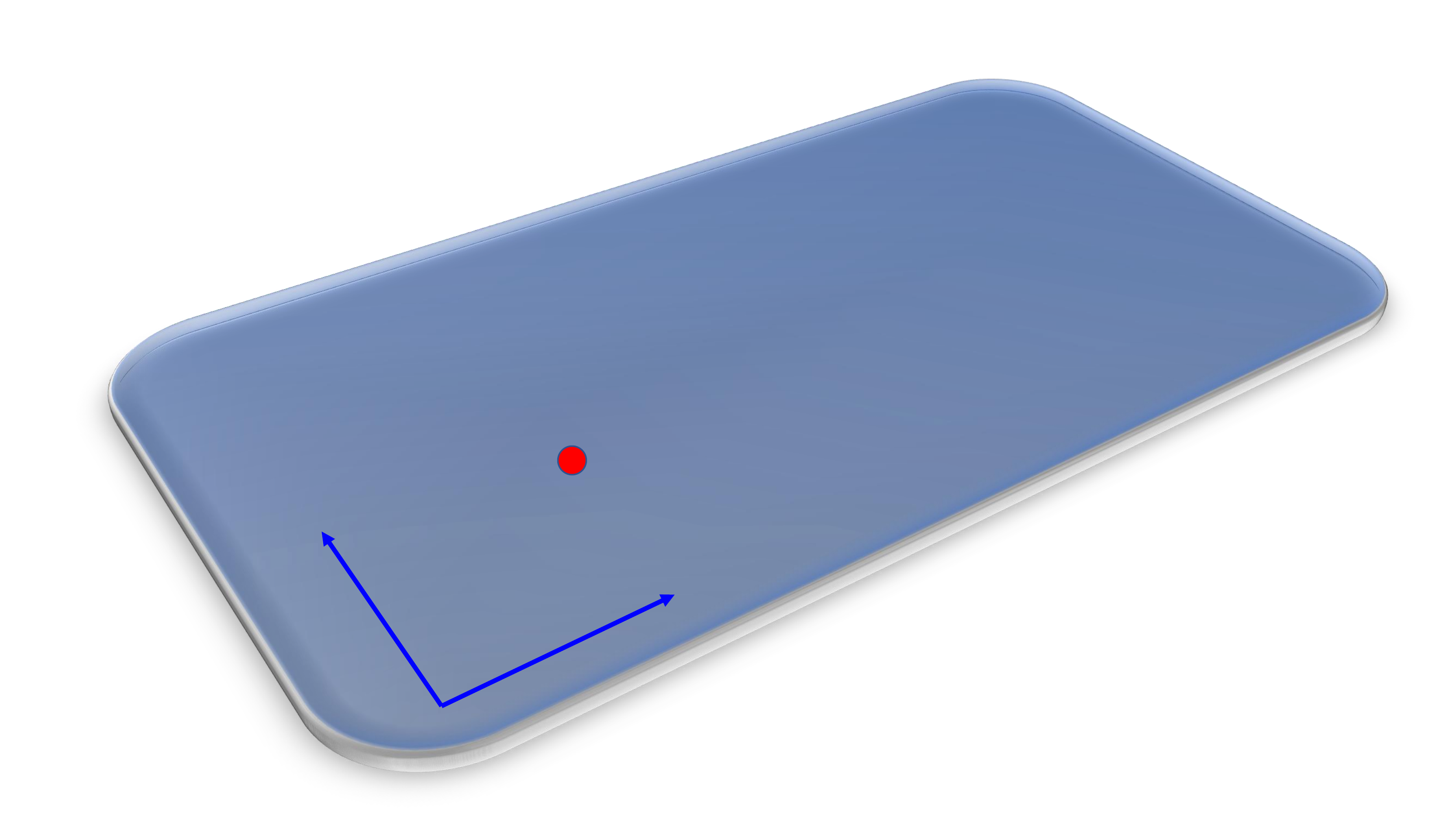}
\put(-145,100){\makebox(0,0)[lb]{\color{yellow}\small$V(x,y)=V_1(x)+V_2(y)$}}
\put(-169,74){\makebox(0,0)[lb]{\color{yellow}\small$m$}}
\put(-222,51){\makebox(0,0)[lb]{\color{blue}\small$y$}}
\put(-142,40){\makebox(0,0)[lb]{\color{blue}\small$x$}}
\caption{The Hamiltonian (\ref{Hcart}) describes a particle with unit mass $m=1$ moving in a two-dimensional potential $V(x,y)=V_1(x)+V_2(x)$. }
\end{figure}
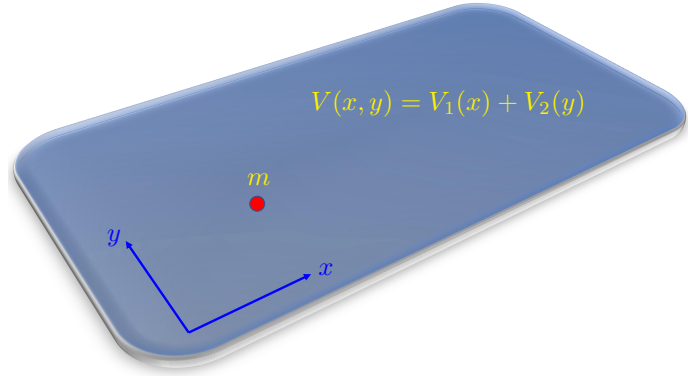

The outline of the paper is as follows. In Section \ref{integralN}, for an arbitrary potential $V(x,y)$ not necessarily separable in a coordinate system we revisited the so called determining equations governing the existence of a general $N$th-order polynomial integral of motion~${\mathcal Y}_N$. In particular, the dominant $N$th-order terms in~${\mathcal Y}_N$ lie in the enveloping algebra of the Euclidean Lie algebra~$e(2)$. From the next leading terms in~${\mathcal Y}_N$, a linear compatibility condition (LCC) can be obtained for the potential~$V$ only. The case of a separable potential in Cartesian coordinates is then analyzed in Sections~\ref{PSCC}--\ref{coefj4}, where we show and describe a~\textit{well} of determining equations and derive the first non-linear compatibility condition for the potential alone. The general form of the potentials is determined by solving these compatibility conditions. Afterwards, the surviving determining equations become linear and can be solved. In Section~\ref{families}, based on the LCC, we~introduce the doubly exotic potentials. A general formula for the corresponding integral~${\mathcal Y}_N$ is given. In Section~\ref{ODEvsAE} we discuss the role of the algebra of the integrals of motion in the search of superintegrable potentials, and a formulation of inverse problem in superintegrability is commented. Section~\ref{N3N4} is devoted to the known examples with $N=3,4$. Finally, in Sections~\ref{Ne5} and~\ref{Rne5} we consider the case $N=5$ and derive in detail all possible doubly exotic potentials. For conclusions see Section~\ref{conclu}.

\section[Superintegrability: existence of an Nth-order polynomial integral]
{Superintegrability: existence of an $\boldsymbol{N}$th-order polynomial\\ integral}\label{integralN}

In the present article we are considering Hamiltonian systems separable in Cartesian coordinates and hence they are second order integrable by construction.
To further search for the superintegrability, we need to give the conditions for the existence of an additional integral of motion, which is a polynomial of order $N>2$ in variables $p_1$, $p_2$. Although
the general ideas for the existence of a $N$th-order integral is given in~\cite{Hietarinta1987, PostWinternitz:2015}, here we would like to summarize those results for the sake of completeness.

\subsection{General form}

The most general form of an $N$th-order polynomial integral ${\mathcal Y}_N$ is given by
\begin{gather}\label{YNQSd}
{\mathcal Y}_N = \sum_{\ell=0}^{[\frac{N}{2}]}\sum_{j=0}^{N-2\ell} f_{j,2\ell}\, p_1^j p_2^{N-j-2\ell},
\end{gather}
see \cite{Hietarinta1987, PostWinternitz:2015}, where ${f}_{j,2\ell}={f}_{j,2\ell}(x,y,V)$ are assumed to be real functions which depend on the coordinates $x$ and $y$ and the potential $V(x,y)$.

The integral ${\mathcal Y}_N$ (\ref{YNQSd}) can be conveniently rewritten as follows
\begin{gather}\label{Yq}
{\mathcal Y}_N = W_N + \text{lower order terms},
\end{gather}
where the leading term $W_{N}$ in (\ref{Yq})
\begin{gather}\label{YNq}
W_{N} = \sum_{0\leq m+n\leq N} A_{N-m-n,m,n} L_z^{N-m-n} p_1^m p_2^n,
\end{gather}
plays a fundamental role since it governs the existence or non-existence of the integral ${\mathcal Y}_N$, here $A_{N-m-n,m,n}$ are $\frac{(N+1)(N+2)}{2}$ real parameters and $L_z=x p_2-y p_1$ is the $z$-component of angular momentum. {If the quantity~${\mathcal Y}_N$ Poisson commutes with the Hamiltonian (\ref{Hcart}) then the system becomes $N$th-order ($N>1$) \emph{superintegrable}. In fact, for 2D systems it would correspond to maximal superintegrability.}

\subsection{The determining equations}

The Poisson bracket of ${\mathcal Y}_N$ (\ref{YNQSd}) with the Hamiltonian $\mathcal H$ (\ref{Hcart}) gives a polynomial, in $p_1$ and $p_2$, of degree $(N+1)$. Explicitly, we have
\begin{gather*}%\label{HYC}
\{{\mathcal H},{\mathcal Y}_N\}_{\rm PB} = \sum_{n_1+n_2=0}^{N+1} M_{n_1,n_2}p_1^{n_1}p_2^{n_2},
\end{gather*}
where the coefficients $ {M}_{n_1,n_2}= {M}_{n_1,n_2}(x,y;f_{j,2\ell},V,N)$ depend on the variables $x$, $y$, the functions $f_{j,2\ell}$ appearing in the integral ${\mathcal Y}_N$, the potential $V(x,y)$ we are looking for, and they also carry an $N$-dependence. Superintegrability requires
\begin{gather}\label{Zxy}
{M}_{n_1,n_2} = 0, \qquad n_1+n_2=0,1,2,\dots,(N+1),
\end{gather}
($\{{\mathcal H},{\mathcal Y}_N \}=0$). For an arbitrary potential $V(x,y)$ not necessarily separable, the system (\ref{Zxy}) is equivalent to the following set of \textit{determining equations} (DE):
\begin{gather}\label{quant deteq}
(\partial_x{ f}_{j-1,2\ell} + \partial_y{f}_{j,2\ell})
- \big[(j+1){f}_{j+1, 2\ell-2} \big]\partial_x V - \big[(N-2\ell+2-j){f}_{j, 2\ell-2} \big]\partial_y V = 0,
\end{gather}
$\ell=0,1,2,\dots,\big[\frac{N}{2}\big]$, $j=0,1,2,\dots,(N-2\ell)$. In (\ref{quant deteq}), the real functions ${f}_{j, \ell} \equiv 0$ identically for $\ell<0$ and $j < 0$ as well as for $j > N - 2\ell$ (further details can be found in~\cite{Hietarinta1987, PostWinternitz:2015}). The DE correspond to the vanishing of all the coefficients, in the Poisson bracket $\{{\mathcal H},{\mathcal Y}_N\}_{\rm PB}$, multiplying the momentum terms of order $n_1+n_2=k=N+1,N-1,N-3,\dots,(N+1-2\ell)$. In particular, for odd $N$ the coefficient multiplying the zero order term is simply $
 f_{1,N-1}V_1' + f_{0,N-1}V_2' = 0,$ obtained from~(\ref{quant deteq}) by making the replacement $\ell \rightarrow \ell +1$.
The DE govern the existence of the integral ${\mathcal Y}_N$. In general, the system (\ref{quant deteq}) is overdetermined. If the potential $V(x,y)$ is not known a priori, then it must be calculated from the compatibility conditions of the DE.

The structure of the DE (\ref{quant deteq}) can be summarized as follows:
\begin{itemize}\itemsep=0pt
 \item The set of DE (\ref{quant deteq}) can be seen as a \textit{well} of recursive equations. The coefficients $f_{j,2\ell}$ in~${\mathcal Y}_N$ depend on the preceding $f_{j,2k}$, $0\leq k<\ell$.
 \item The bottom level of equations (\ref{quant deteq}) corresponds to $\ell=0$. The associated DE do not depend on $V$, thus, allowing exact solvability. Indeed, they define the coefficient-functions~$f_{j,0}$, $j=0,1,2,\dots,N$. The explicit expression for $f_{j,0}$ is given by
\begin{gather*}
%\label{fj0 Nthorder}
f_{j,0} = \sum_{n=0}^{N-j} \sum_{m=0}^{j}\binom{ N-n-m}{j-m}A_{N-n-m,m,n}x^{N-j-n}(-y)^{j-m},
\end{gather*}
see~\cite{Hietarinta1987, PostWinternitz:2015}. Accordingly, the leading part~(\ref{YNq}) of ${\mathcal Y}_N$ is a polynomial of order $N$ in
the enveloping algebra of the Euclidean Lie algebra $e(2)$ with basis $\{p_1, p_2, L_z\}$.

\item The 2nd level of DE (\ref{quant deteq}) occurs at $\ell=1$. They provide a \emph{linear compatibility condition} (LCC) for the potential $V$ only. For arbitrary potential, this linear PDE can be written in the compact form \cite{Hietarinta1987,PostWinternitz:2015}
\begin{gather}\label{LCC}
\sum_{j=0}^{N-1}{(-1)}^j\partial_x^{N-1-j}\partial_y^{j}\big[ (j+1)f_{j+1,0}\partial_x V + (N-j)f_{j,0}\partial_yV \big] = 0.
\end{gather}
This above equation is a necessary but not sufficient condition for $\{{\mathcal H},{\mathcal Y}_N \}=0$. Also, in the quantum case the LCC remains identical to (\ref{LCC}). However, the corresponding DE do acquire $\hbar$-dependent terms.
 \item Beginning from $\ell=2$, the DE (\ref{quant deteq}) will lead to \emph{nonlinear compatibility conditions} (NLCC) for the potential $V$ alone. We~should remind here that in the quantum case these NLCC, unlike the LCC, do depend non-trivially on $\hbar$. Hence, the classical and quantum cases can greatly differ, and it requires to treat them separately.
\end{itemize}

\section{Superintegrable potentials separable in Cartesian coordinates}\label{PSCC}

\subsection{The linear compatibility condition}

In the case of a separable potential the LCC (\ref{LCC}) leads to the ordinary differential equations for $V_1(x)$
\begin{gather}\label{QX}
\sum_{j=0}^{N-1}(j+1)! \sum_{n=0}^{N-j-1}\binom{N-1-n}{j} A_{N-1-n,1,n}\bigg(\frac{\rm d}{{\rm d}x}\bigg)^{N-j+1} \big[x^{N-j-n-1} V_1'(x)\big] = 0,
\end{gather}
and
\begin{gather}\label{QX2}
 \sum_{j=0}^{N-1}(j\!+1)(j\!+1)! (-1)^{2j+1} \!\sum_{n=0}^{N-j-1}\!\!\binom{N\!-n}{j\!+1} A_{N-n,0,n}\bigg(\frac{\rm d}{{\rm d}x}\bigg)^{N-j+1}\!\! \big[x^{N-j-n-1} V_1'(x)\big]\! = 0.
\end{gather}
For superintegrability, $\{{\mathcal H},{\mathcal Y}_N \}=0$, these two linear equations (\ref{QX}) and~(\ref{QX2}) must be simultaneously satisfied. Similarly, for $V_2(y)$ there exist two ODEs which can be obtained from (\ref{QX}) and~(\ref{QX2}) using the symmetry $x\leftrightarrow y$, respectively.

\section{The first nonlinear compatibility condition}\label{coefj4}

In the case of an arbitrary odd {$N \geq 3$} polynomial integral of motion ${\mathcal Y}_N$, following the derivation presented in~\cite{AMGen} we describe the procedure to construct the first NLCC in detail. In general, this equation obtained from the DE (\ref{quant deteq}) with $\ell=2$ provides the form of the doubly exotic potentials, see below.

As a first step, one solve the DE (\ref{quant deteq}) with $\ell=1$. These equations define all the coefficient-functions ${f}_{j,2}$ appearing in the integral (\ref{YNQSd}).

Secondly, from the DE (\ref{quant deteq}) with $\ell=2$ we compute the ($N-3$) functions ${f}_{j,4}$ except those with $j=\frac{N-5}{2}$ and $j=\frac{N-3}{2}$. Eventually, we arrive at the equations
\begin{gather}\label{NLCCNODD}
\partial_y{f}_{\frac{N-5}{2},4} = \tilde F_\frac{N-5}{2},\qquad
\partial_x{ f}_{\frac{N-5}{2},4} + \partial_y{ f}_{\frac{N-3}{2},4} = \tilde F_\frac{N-3}{2},\qquad
\partial_x{ f}_{\frac{N-3}{2},4} = \tilde F_\frac{N-1}{2},
\end{gather}
here the $\tilde F$'s, by construction, are real functions that solely depend on the potential $V$ (and its derivatives).

Finally, from (\ref{NLCCNODD}) it follows the equation
\begin{gather*}%\label{Noddg}
\partial^2_{x}\tilde F_\frac{N-5}{2} + \partial^2_{y}\tilde F_\frac{N-1}{2} - \partial^2_{x,y}\tilde F_\frac{N-3}{2} \equiv 0,
\end{gather*}
which gives the aforementioned NLCC for the potential $V$. In the case of arbitrary even {$N \geq 4$}, the steps are quite similar, see details in~\cite{AMGen}.

From (\ref{quant deteq}), it follows that more NLCC occur for each value of $\ell=3,4,\dots,\big[\frac{N}{2} \big]$. Nevertheless, these additional equations will simply restrict the general solution of the potential $V$ found from the previous NLCC with $\ell=2$.

Therefore, for a separable potential $V=V_1(x)+V_2(y)$ the set of DE with $\ell=0$ are given by a system of ODEs which do not depend on $V$ and they specify the coefficient-functions $f_{j,0}$ ($j=0,1,2,\dots,N$) appearing in (\ref{YNQSd}). Then, the next level of DE with $\ell=1$ provide a LCC for the potential alone and they also determine the functions $f_{j,2}$ ($j=0,1,2,\dots,N-2$). At all further levels $\ell\geqslant 2$ the DE and their compatibility conditions are nonlinear ODEs for $V$. These compatibility
conditions are instrumental to specify the general form of the potential $V$.

\section{Doubly exotic potentials}\label{families}

Hereafter, we will restrict ourselves to the case of doubly exotic potentials. These potentials satisfy the LCC (\ref{LCC}) trivially. In particular, the two linear ODEs (\ref{QX}) and~(\ref{QX2}) vanish identically for any $V_1(x)$. Hence, this LCC does not impose any constraint for $V_1(x)$ nor for $V_2(y)$. This situation occurs when the number of coefficients $A_{N-m-n,m,n}$ that figure in the LCC is less that those appearing in the integral $Y_N$ (\ref{YNQSd}). In this case, we simply put equal to zero the coefficients $A_{N-m-n,m,n}$ in the LCC, thus, it vanishes identically, but still the integral $Y_N$ is of order $N$. In general, based on the LCC (\ref{LCC}) one can classify the $N$th-order superintegrable systems into three major classes: doubly exotic potentials, singly exotic potentials and standard potentials (see~\cite{AMGen}). This general classification is summarized in Table~\ref{FiC}.
\begin{table}[h]%\setlength{\tabcolsep}{5.5pt}
\caption{Classification of $N$th-order superintegrable classical systems ($N>2$) separating in Cartesian coordinates. For a fixed value of $N$, there exist three generic types of potentials: doubly standard, doubly exotic and singly exotic potentials.}\label{FiC}\vspace{4pt}\centering
{\small\begin{tabular}{c|c|c|c}
\hline
Potential\rule{0pt}{12pt} & Doubly standart & Doubly exotic &
Singly exotic
\\
$V=V_1(x)+V_2(y)$ &potentials &potentials &potentials
\\[2pt]
\hline
Classical\rule{0pt}{12pt}
&Both functions
&Both $V_1(x)$, $V_2(y)$
&The $x$-component
\\
superintegrable &$V_1(x)$, $V_2(y)$ satisfy
&obey a NLCC,
& $V_1(x)$ satisfies
\\
systems & non-trivially
&a non-linear ODE
&a linear/non-linear
\\
&the LCC, &which do not pass
&ODE whilst $y$-component
\\
&a linear ODE&the Painlev\'{e} test. &$V_2(y)$ obeys
\\
&&The LCC is identically zero&a non-linear/linear OD
\\[2pt]
\hline
\end{tabular}}
\end{table}

From this point of view, Cases 1--3 of Proposition 1 presented in~\cite{Grigoriev} are doubly exotic potentials for $n_1,n_2>1$ whilst Cases 4--5 at $n>1$ can not be doubly standard ones. Moreover, the aforementioned Cases 1--3 are nothing but particular solutions of the present direct approach. It is worth mentioning that we solely consider potentials where neither the $x$-part $V_1(x)$ nor the $y$-part $V_2(y)$ are constant functions.

\subsection[Integral \protect{$Y\_N$} for doubly exotic potentials]
{Integral $\boldsymbol{{\mathcal Y}_N}$ for doubly exotic potentials}

In the present work we will focus on doubly exotic potentials. In this case, the corresponding $N$th-order terms of the integral ${\mathcal Y}_{N}$ (\ref{YNQSd}) are given by
\begin{align}
W_N = {}&A_{0,N,0}p_1^N + A_{0,0,N}p_2^N + A_{N-4,2,2} L_z^{N-4} p_1^2 p_2^2\nonumber
\\
& + \sum_{4 < m+n < N;\,| m-n| < N-4} A_{N-m-n,m,n}L_z^{N-m-n} p_1^mp_2^n \nonumber
\\
&+ \sum_{0 \leq m+n = N;\,| m-n| \leq N-4} A_{0,m,n} p_1^mp_2^n.
\label{PNQ}
\end{align}
Therefore, $W_N$ (\ref{PNQ}) carries $\frac{1}{2}\big(10-5N+N^2\big)$ less constants $A_{N-m-n,m,n}$ than the generic term~(\ref{YNQSd}). For large $N$, the number of constants $A_{N-m-n,m,n}$ in $W_N$ grows quadratically with~$N$. Notice that $L_z$ occurs in $W_N$ starting from $N=5$. For $N\geq5$, it can also contain terms like $L_z, L_z^2, L_z^3,\dots,L_z^{N-4}$ only.

Let us give the most general (leading) term $W_N$ of the integral ${\mathcal Y}_{N,{\rm doubly\ exotic}}$ for $N=3,4,5$ explicitly
\begin{gather*}%\label{YN3de}
W_3 = A_{030}p_1^3 + A_{003}p_2^3,% \nonumber
\\%\label{YN4de}
W_4 = A_{040}p_1^4 + A_{004}p_2^4 + A_{022}p_1^2 p_2^2, %\nonumber
\\%\label{YN5de}
W_5 = A_{050}p_1^5 + A_{005}p_2^5 + A_{032}p_1^3p_2^2 + A_{023}p_1^2p_2^3 + A_{122}L_zp_1^2 p_2^2. %\nonumber
\end{gather*}

\section[ODEs versus algebraic equations. Algebras of integrals of motion]
{ODEs versus algebraic equations. Algebras of integrals \\of motion}\label{ODEvsAE}

In the search of $N$th-order superintegrable potentials one faces the problem of solving an overdetermined system of ODEs where some of them are non-linear. Moreover, the number of involved equations increases with $N$. Therefore, the direct approach of solving all the DE (\ref{quant deteq}) is far from being an efficient method. In order to simplify it, in the present consideration we propose to combine two basic elements, namely the non-linear compatibility conditions and the use of the algebra of the integrals (see below). As a result, in some cases the ODEs are reduced to pure algebraic equations.

From $\mathcal X$ and ${\mathcal Y}_N$, we introduce the quantity
\begin{gather}\label{PIC}
 C \equiv \{{\mathcal Y}_N,{\mathcal X}\}_{\rm PB},
\end{gather}
which is a polynomial function in $p_1$ and $p_2$ of degree ($N+1$). If ${\mathcal Y}_N$ is an integral of motion, then by construction $C$ (\ref{PIC}) is also conserved. The closure of the algebra generated by the integrals of motion $({\mathcal H},{\mathcal X},{\mathcal Y}_N,C)$ is guaranteed by the property of maximal superintegrability. The main question we aim to explore is the appearance and utility of a \emph{closed polynomial algebra}.

It is important to mention that the study of the algebraic structure of the integrals of motion has been proven to be fruitful in the classification of higher-order superintegrable classical and quantum systems \cite{Daskol,IanM}.

Also, the explicit results obtained in Section \ref{Rne5} suggests to explore the inverse problem, namely we take two polynomial functions ${\mathcal A}$ and ${\mathcal B}$ in momentum variables $(p_1,p_2)$ and construct the new object ${\mathcal C}=\{{\mathcal A},{\mathcal B}\}_{\rm PB}.$ Now, let us assume that the algebra generated by $({\mathcal A},{\mathcal B},{\mathcal C})$ is a closed polynomial algebra with polynomial coefficients in ${\mathcal H}
$. The question is \textit{under what conditions these closure relations imply that ${\mathcal A}$ and ${\mathcal B}$ are integrals, i.e., they Poisson commute with $\mathcal H$}?

\section[Lowest order cases N=3 and N=4: doubly exotic potentials]
{Lowest order cases $\boldsymbol{N=3}$ and $\boldsymbol{N=4}$: \\doubly exotic potentials}\label{N3N4}

\subsection[Case N=3]
{Case $\boldsymbol{N=3}$}

The general integral (\ref{YNQSd}) at $N=3$ is given by
\begin{gather}\label{Y3ggene}
{\mathcal Y}_3 = f_{3,0}p_1^3 + f_{0,0}p_2^3 + {f}_{1,0} p_1p_2^2 + {f}_{2,0} p_2p_1^2 + {f}_{1,2} p_1 + {f}_{0,2} p_2.
\end{gather}

The first set of DE (\ref{quant deteq}) with $\ell=0$ corresponds to the vanishing of all the coefficients, in the Poisson bracket $\{{\mathcal H},{\mathcal Y}_3\}_{\rm PB}$, multiplying the highest momentum terms of order $4$. They can be solved directly to give the functions $f_{j,0}$. For doubly exotic potentials, they read
\begin{gather*}
 f_{30} = A_{030},
\qquad
f_{20} = 0,
\qquad
f_{10} = 0,
\qquad
f_{00} = A_{003}.
\end{gather*}
Thus, (\ref{Y3ggene}) reduces to
\begin{gather}\label{IYEn3}
{\mathcal Y}_3 = A_{030}p_1^3 + A_{003}p_2^3 + {f}_{1,2} p_1 + {f}_{0,2} p_2.
\end{gather}
The next set of DE is obtained by setting $\ell=1$ in (\ref{quant deteq}). They correspond to the vanishing of all the coefficients, in the Poisson bracket $\{{\mathcal H},{\mathcal Y}_3\}_{\rm PB}$, multiplying the (next-to-leading) momentum terms of order $2$. These DE take the form
\begin{gather}
 f_{1,2}^{(1,0)} = 3A_{030}V_1',%\nonumber
\qquad
f_{1,2}^{(0,1)} + f_{0,2}^{(1,0)} = 0,%\nonumber
\qquad
f_{0,2}^{(0,1)} = 3A_{003}V_2'.
\label{N3f}
\end{gather}
The compatibility condition of the above system (\ref{N3f}) does not provide further information on the potentials functions. However, the first and third equations can be solved immediately, they define the functions $f_{0,2}$ and $f_{1,2}$
\begin{gather}
f_{1,2} = 3A_{030}V_1 + u_2(y),%\nonumber
\qquad
f_{0,2} = 3A_{003}V_2 + u_1(x),
\label{N3fs}
\end{gather}
where $u_1(x)$ and $u_2(y)$ are arbitrary functions of $x$ and $y$, respectively. Substituting (\ref{N3fs}) into the second equation in (\ref{N3f}) we obtain the equation $u_1' + u_2' = 0$. Therefore,
\begin{gather*}
u_1 = \alpha_1 + \beta x, \qquad u_2 = \alpha_2 - \beta y,
\end{gather*}
here $\alpha_1$, $\alpha_2$ are constants of integrations whilst $\beta$ is a separation constant. Finally, the last determining equation corresponds to the vanishing of the coefficient, in the Poisson bracket $\{{\mathcal H},{\mathcal Y}_3\}_{\rm PB}$, of order zero in momentum variables. Explicitly, it takes the form
\begin{gather}
\label{N3DM0}
3 A_{030} V_1 V_1' + 3 A_{003} V_2 V_2' + \alpha_2 V_1' - \beta y V_1' + \beta x V_2' + \alpha_1 V_2' = 0.
\end{gather}
Non trivial solutions of (\ref{N3DM0}) correspond to separation of variables, namely $\beta=0$. In this case, (\ref{N3DM0}) leads to the following uncoupled equations
\begin{gather*}
 3 A_{030} V_1 V_1' + \alpha_2 V_1' = \lambda,
\qquad
3 A_{003} V_2 V_2' + \alpha_1 V_2' = -\lambda,
\end{gather*}
being $\lambda \neq 0$ {(otherwise the functions $V_{1,2}$ are just constants)} the corresponding separation constant. Eventually, we arrive to the solutions
\begin{gather}
V_1 = \sqrt{\frac{2\lambda}{3A_{030}}}\sqrt{x},\qquad
V_2 = \sqrt{\frac{-2\lambda}{3A_{003}}}\sqrt{y}.
\label{N3SOL}
\end{gather}
In general, the Poisson bracket $C = \{{\mathcal Y}_3,{\mathcal X} \}_{\rm PB}$ between ${\mathcal Y}_N$ (\ref{YNQSd}) with $N=3$ and $\mathcal X$ (\ref{X}) is a~polynomial in the variables $p_1$ and $p_2$ of degree four. However, for the potential functions (\ref{N3SOL}), we obtain $C \propto \lambda $. Hence, in this case the algebra of the three integrals of motion ($C,{\mathcal Y}_3,{\mathcal X}$) takes the form
\begin{gather*}
\{ C,{ \mathcal X}\}_{\rm PB} = 0, \qquad
\{ C,{\mathcal Y}_3 \}_{\rm PB} = 0.
\end{gather*}

\begin{figure}[h]\centering\vspace{2ex}
\includegraphics[scale=0.4]{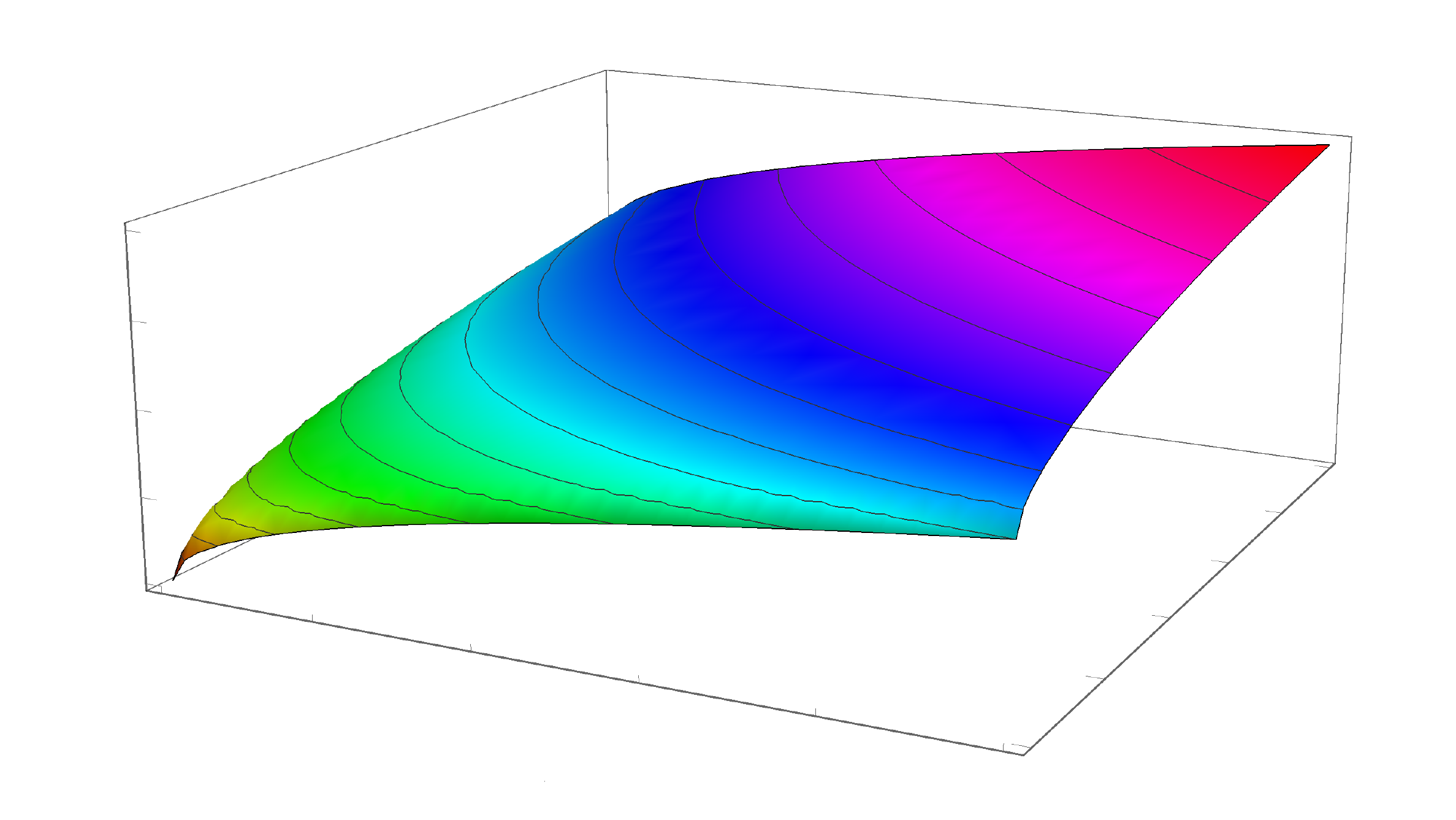}
\put(-170,135){\makebox(0,0)[lb]{\small$V=V_1(x)+V_2(y)$}}
\put(-236,30){\makebox(0,0)[lb]{\small$0$}}
\put(-245,46){\makebox(0,0)[lb]{\small$0.5$}}
\put(-246,63){\makebox(0,0)[lb]{\small$1.0$}}
\put(-247,80){\makebox(0,0)[lb]{\small$1.5$}}
\put(-248,98){\makebox(0,0)[lb]{\small$2.0$}}
\put(-228,22){\makebox(0,0)[lb]{\small$0$}}
\put(-205,16){\makebox(0,0)[lb]{\small$0.2$}}
\put(-175,11){\makebox(0,0)[lb]{\small$0.4$}}
\put(-155,0){\makebox(0,0)[lb]{\small$x$}}
\put(-143,5){\makebox(0,0)[lb]{\small$0.6$}}
\put(-110,-1){\makebox(0,0)[lb]{\small$0.8$}}
\put(-74,-8){\makebox(0,0)[lb]{\small$1.0$}}
\put(-57,-2){\makebox(0,0)[lb]{\small$0$}}
\put(-45,10){\makebox(0,0)[lb]{\small$0.2$}}
\put(-32,22){\makebox(0,0)[lb]{\small$0.4$}}
\put(-4,23){\makebox(0,0)[lb]{\small$y$}}
\put(-21,32){\makebox(0,0)[lb]{\small$0.6$}}
\put(-10,42){\makebox(0,0)[lb]{\small$0.8$}}
\put(-2,52){\makebox(0,0)[lb]{\small$1.0$}}
\caption{The doubly exotic potential (\ref{N3SOL}) corresponding to $N=3$. It admits the third-order integ\-ral~${\mathcal Y}_3$~(\ref{IYEn3}). The values $A_{030}=-A_{003}=\frac{2\lambda}{3}$ were used.}
%\label{}
\end{figure}
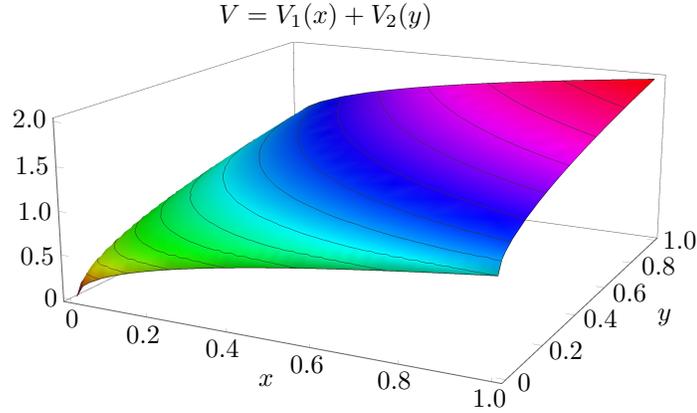

It is worth mentioning that the family of superintegrable potentials with
\[
{\mathcal Y}_N = A_{0N0}p_1^N + A_{00N}p_2^N + {\rm lower\ order\ terms},
\]
has been analyzed in~\cite{GungorKNN:2014} by means of {H}eisenberg-type higher order symmetries. For this family all three conserved quantities $({\mathcal H},{\mathcal X},{\mathcal Y}_N)$ admit separation of variables in Cartesian coordinates. However, such an approach does not allow us to obtain all the doubly exotic potentials with non-separable integrals ${\mathcal Y}_N$.

\subsection[Case N=4]
{Case $\boldsymbol{N=4}$}

In this case $N=4$, for a doubly exotic potential the most general expression of the fourth-order integral ${\mathcal Y}_4$ reads
\begin{gather*}
{\mathcal Y}_4 = A_{040}p_1^4 + A_{004}p_2^4 + A_{022}p_1^2 p_2^2 + {\rm lower\ order\ terms},
\end{gather*}
where $A_{040}$, $A_{004}$ and $A_{022}$ are real constants. It can immediately be rewritten as follows
\begin{gather}\label{Y4DEinte}
{\mathcal Y}_4 = A_{040}{( {\mathcal H}+{\mathcal X} )}^2 + A_{004}{( {\mathcal H}-{\mathcal X} )}^2 + A_{022}{( {\mathcal H}+{\mathcal X} )} {( {\mathcal H}-{\mathcal X} )} + {\rm lower\ order\ terms}.
\end{gather}
Now, without losing generality, one can always add to (\ref{Y4DEinte}) any arbitrary function of the second order trivial integrals ${\mathcal H}$ and $\mathcal X$. This implies that no \emph{bona fide} doubly exotic potentials, with a~non-trivial fourth order integral, exist.

\section[Case N=5: doubly exotic potentials]
{Case $\boldsymbol{N=5}$: doubly exotic potentials}\label{Ne5}

We can write the most general $5$th-order polynomial integral ${\mathcal Y}_5$ in the form
\begin{gather}\label{Y5QSd}
{\mathcal Y}_5 = \sum_{\ell=0}^{2}\sum_{j=0}^{5-2\ell} f_{j,2\ell}\, p_1^j p_2^{5-j-2\ell}.
\end{gather}

\subsection{Determining equations}

Putting $\ell=0$ in (\ref{quant deteq}) corresponds to the vanishing of all the coefficients, in the Poisson bracket $\{{\mathcal H},{\mathcal Y}_5\}_{\rm PB}$, multiplying the highest momentum terms of order $6$. They can be solved directly to give the functions $f_{j,0}$
\begin{gather*}
 f_{50} = A_{050},
\qquad
f_{40} = 0,
\qquad
f_{30} = A_{032} - y A_{122},
\\
f_{20} = A_{023} + x A_{122},
\qquad
f_{10} = 0,
\qquad
f_{00} = A_{005},
\end{gather*}
where the condition that the LCC (\ref{LCC}) is satisfied trivially was imposed, namely we consider doubly exotic potentials only. It implies that the existence or non-existence of fifth order doubly\allowdisplaybreaks exotic potentials is governed by 5 parameters $A_{050}$, $A_{005}$, $A_{032}$, $A_{023}$, $A_{122}$ only. The next set of~DE are obtained by setting $\ell=1$:
\begin{gather}
 f_{0,2}{}^{(0,1)} = 5 A_{005} V_2',\nonumber
\\
 f_{1,2}{}^{(0,1)} + f_{0,2}{}^{(1,0)} = 2 ( A_{023} + x A_{122} ) V_1',\nonumber
\\
 f_{1,2}{}^{(1,0)} + f_{2,2}{}^{(0,1)} = 3 ( A_{032} - y A_{122} ) V_1' + 3 ( A_{023} + x A_{122} ) V_2',\nonumber
\\
 f_{3,2}{}^{(0,1)} + f_{2,2}{}^{(1,0)} = 2 ( A_{032} - y A_{122} ) V_2',\nonumber
\\
f_{3,2}{}^{(1,0)} = 5 A_{050} V_1'.
\label{D1}
\end{gather}
Now, the three DE (\ref{quant deteq}) with $\ell=2$ are given by
\begin{gather}
 f_{1,4}{}^{(1,0)} = 3 f_{3,2} V_1' + f_{2,2} V_2',\nonumber
\\
 f_{1,4}{}^{(0,1)} + f_{0,4}{}^{(1,0)} = 2 (f_{2,2} V_1' + f_{1,2} V_2' ),\nonumber
\\
 f_{0,4}{}^{(0,1)} = 3 f_{0,2} V_2' + f_{1,2} V_1'.
\label{D2}
\end{gather}
Next, following the discussion of Section \ref{coefj4} we obtain from (\ref{D1}) the functions $f_{3,2}$, $f_{2,2}$, $f_{1,2}$, $f_{0,2}$ in terms of $V$ (see below). Afterwards, the r.h.s.\ in (\ref{D2}) would depend (non-linearly) on $V$ and its derivatives alone. Consequently, (\ref{D2}) leads to the first NLCC in the form
\begin{gather}\label{D3}
 \partial^2_x f_{0,4}^{(0,1)} + \partial^2_y f_{1,4}^{(1,0)} - \partial_x \partial_y \big(f_{0,4}^{(1,0)} - f_{1,4}^{(0,1)}\big) = 0.
\end{gather}
Finally, the last determining equation with {$\ell=2$} reads
\begin{gather*}
 f_{1,4}V_1' + f_{0,4}V_2' = 0.
\end{gather*}

\subsection{The (first) NLCC}

The DE with $\ell=1$ (\ref{D1}) define the four functions $f_{0,2}$, $f_{1,2}$, $f_{2,2}$ and $f_{3,2}$ appearing in the integral~${\mathcal Y}_5$ (\ref{Y5QSd}) in front of the cubic terms ($p_1^{i} p_2^{j}$ with $i+j=3$). Explicitly
\begin{gather}
 f_{0,2} = 2 x A_{122} T_1'(x) + 2 A_{023} T_1'(x) + A_{122} T_1(x) + 5 A_{005} T_2'(y) + \alpha_1 - \beta_4 x^3 + \sigma_3 x^2 + \alpha_2 x,\nonumber
\\
f_{1,2} = y ( -3 A_{122} T_1'(x) - \alpha _2+3 \beta _4 x^2-2 \sigma _3 x ) + 3 A_{032} T_1'(x)
+ \nu _1 + \nu _3 x^2 - \sigma _2 x,\nonumber
\\
f_{2,2} = x ( 3 A_{122} T_2'(y)-\beta _2-3 \beta _4 y^2-2 \nu _3 y ) + 3 A_{023} T_2'(y) + \sigma _1 + \sigma _3 y^2 + \sigma_2 y,\nonumber
\\
f_{3,2} = -2 y A_{122} T_2'(y) + 2 A_{032} T_2'(y) - A_{122} T_2(y) + 5 A_{050} T_1'(x)\nonumber
\\ \hphantom{f_{3,2} =}
{} + \beta _1 + \beta _4 y^3 + \nu _3 y^2 + \beta _2 y,
\label{fsN5de}
\end{gather}
where
\begin{gather*}
V_1(x) \equiv T_1'(x), \qquad V_2(y) \equiv T_2'(y).
\end{gather*}
Next, substituting (\ref{D2}) and (\ref{fsN5de}) into (\ref{D3}) we obtain the following non-linear compatibility condition (NLCC)
\begin{gather}
\text{NLCC} = T_1{}^{(4)} (3 T_1' (A_{032}-y A_{122})+\nu _1+3 \beta _4 x^2 y+x (-\sigma _2+\nu _3 x-2 \sigma _3 y)-\alpha _2 y )\nonumber
\\ \hphantom{\text{NLCC} =}
{} + T_2{}^{(4)} (3 T_2' (x A_{122}+A_{023})+\sigma _1-3 \beta _4 x y^2+y (\sigma _2+\sigma _3 y -2 \nu _3 x )-\beta _2 x)\nonumber
\\ \hphantom{\text{NLCC} =}
 {}+ T_1{}^{(3)}(-9 y A_{122} T_1''+9 A_{032} T_1''-4 \sigma _2+8\nu _3 x+24 \beta _4 x y-8 \sigma _3 y)\nonumber
\\ \hphantom{\text{NLCC} =}
{} + T_2{}^{(3)} (9 x A_{122} T_2''+9 A_{023} T_2''+4 \sigma _2-8 \nu _3 x-24 \beta _4 x y+8 \sigma _3 y)\nonumber
\\ \hphantom{\text{NLCC} =}
{} + 12\nu _3 T_1'' + 12 \sigma _3 T_2'' - 36 \beta _4 x T_2'' + 36 \beta _4 y T_1'' = 0,
 \label{Enlcc}
\end{gather}
where $\alpha_1$, $\beta'$s, $\nu'$s and $\sigma'$s are constants to be determined. We~have the freedom to replace $T_{1(2)}$ by $T_{1(2)}+c $ for some real constant $c$ to simplify the expressions. Also we can shift the variables~$x$ or $y$. {Notice that the constants $A_{050}$ and $A_{005}$ do not appear in (\ref{Enlcc}).}

In terms of the parameters $A_{5-m-n,m,n}$ that define the existence or non-existence of the integral ${\mathcal Y}_5$, we identify two cases for which the above NLCC (\ref{Enlcc}) admits separation of variables in Cartesian coordinates, namely:
\begin{itemize}\itemsep=0pt
 \item[$(i)$] $A_{122} \neq 0$, $A_{023}=A_{032}=0 $,
 \item[$(ii)$] $A_{023}^2 + A_{032}^2 \neq 0$, $A_{122}=0$,
\end{itemize}
with $A_{050}$ and $A_{005}$ arbitrary. These two cases are ${\mathcal{S}}_2$-invariant under the permutation $x \Leftrightarrow y$ (thus, $p_1 \Leftrightarrow p_2$). Let us recall that the Hamiltonian $\mathcal H$ and the integral $\mathcal X$ are ${\mathcal{S}}_2$-invariant and ${\mathcal{S}}_2$-antiinvariant, respectively. Moreover, if
\begin{itemize}\itemsep=0pt
 \item [$(iii)$] $A_{122}= A_{023}= A_{032} = 0,$
\end{itemize}
with $A^2_{050}+A^2_{005}\neq 0$, the NLCC degenerates into a linear equation which must be identically zero for doubly exotic potentials. In such a case the NLCC does not provide any information on the potential.
As a result of calculations, the cases $(i)$, $(ii)$ and $(iii)$ are the only generic ones that satisfy all the DE.

\section{Results}\label{Rne5}

\subsection{Superintegrable potentials}

Below, {adopting the notation introduced in~\cite{AW}} we present the full list of doubly exotic fifth-order ($N=5$) superintegrable potentials:

\medskip
{\it Case} $(i)$.

$\bullet$ The system $Q_1$: $A_{122}\neq 0$, $A_{032}=A_{023}=A_{050}=A_{005}=0$.
This system corresponds to $A_{122}=1$, all other parameters $A_{ijk}=0$. In this case, by solving all the DE (\ref{D1})--(\ref{D3}) we eventually arrive to the first-order nonlinear ODEs
\begin{gather}
(T_1'){}^2 - 2 \beta_4 x^2 T_1' - 4 \beta _4 T_1 x + \beta _4^2 x^4 = 0,\nonumber
\\
(T_2'){}^2 - 2 \beta _4 y^2 T_2' - 4 \beta_4 T_2 y + \beta_4^2 y^4 = 0,
\label{VQ1}
\end{gather}
$\beta_4 \neq 0$ is a real constant. The corresponding fifth-order integral of motion is given by
\begin{gather}
{\mathcal Y}_5^{(Q_1)} = p_1^2 p_2^3 x - p_1^3 p_2^2 y + p_1^3 \big({-}2 y T_2'-T_2+\beta _4 y^3\big) + p_2^3 \big(2 x T_1'+T_1-\beta _4 x^3\big)\nonumber
\\ \hphantom{{\mathcal Y}_5^{(Q_1)} = }
{}+ p_1^2 p_2 x \big(3 T_2'-3 \beta _4 y^2\big) + p_1 p_2^2 y \big(3 \beta _4 x^2-3 T_1'\big) \nonumber
\\ \hphantom{{\mathcal Y}_5^{(Q_1)} = }
{}+p_1 \bigg({-}\frac{3}{2} \beta _4 x^2 y^2 T_2'' + 3 \beta _4 y^3 T_1' + \frac{3}{2} x^2 T_2' T_2'' - 6 y T_1' T_2' - 3 T_2 T_1'\bigg) \nonumber
\\ \hphantom{{\mathcal Y}_5^{(Q_1)} = }
{}+ p_2 \bigg(\frac{3}{2} \beta _4 x^2 y^2 T_1'' - 3 \beta _4 x^3 T_2' - \frac{3}{2} y^2 T_1' T_1'' + 6 x T_1' T_2' + 3 T_1 T_2'\bigg).
\label{Y5Q1}
\end{gather}

From $\mathcal X$ and ${\mathcal Y}_5^{(Q_1)}$, we built the quantity\vspace{-1ex}
\begin{gather*}%\label{CQ1}
 C \equiv \big\{{\mathcal Y}_5^{(Q_1)},{\mathcal X}\big\}_{\rm PB},
\end{gather*}
which is a polynomial function in $p_1$ and $p_2$ of sixth degree. By construction, it is an integral when (\ref{VQ1}) are satisfied. Now, if we demand that the three elements $\big(\mathcal X$, ${\mathcal Y}_5^{(Q_1)},C\big)$ generate a~\emph{closed polynomial algebra} we eventually arrive to a nonlinear first-order differential equation for $T_1(x)$ and similarly for $T_2(y)$. Therefore, from these equations and (\ref{VQ1}) we can eliminate the first-derivative $T_1'$, $T_2'$ terms and obtain
an algebraic equation for both $T_1(x)$ and $T_2(y)$. The solutions of such algebraic equations turn out to be the general solutions of (\ref{VQ1}). Explicitly, these algebraic equations take the form
\begin{gather}
3\beta _4 T_1^2 + 8 \beta _4^{3/2} x^{3/2} T_1^{3/2} + 6 \beta _4^2 x^3 T_1 - \beta _4^3 x^6 - \delta = 0,\nonumber
\\
3 \beta _4 T_2^2 + 8 \beta _4^{3/2} y^{3/2} T_2^{3/2} + 6 \beta _4^2 y^3 T_2 - \beta _4^3 y^6 - \delta = 0,
\label{AEQ1}
\end{gather}
where $\delta$ is an arbitrary constant. In the case $\delta=0$, we immediately obtain the particular solutions
\begin{equation*}
 T_1(x) = \beta_4x^3,\ \frac{\beta_4}{9}x^3, \qquad \text{and} \qquad
 T_2(y) = \beta_4y^3,\ \frac{\beta_4}{9}y^3,
\end{equation*}
which correspond to a well-known lower-order superintegrable system.

\begin{figure}[h]\centering
\includegraphics[scale=0.4]{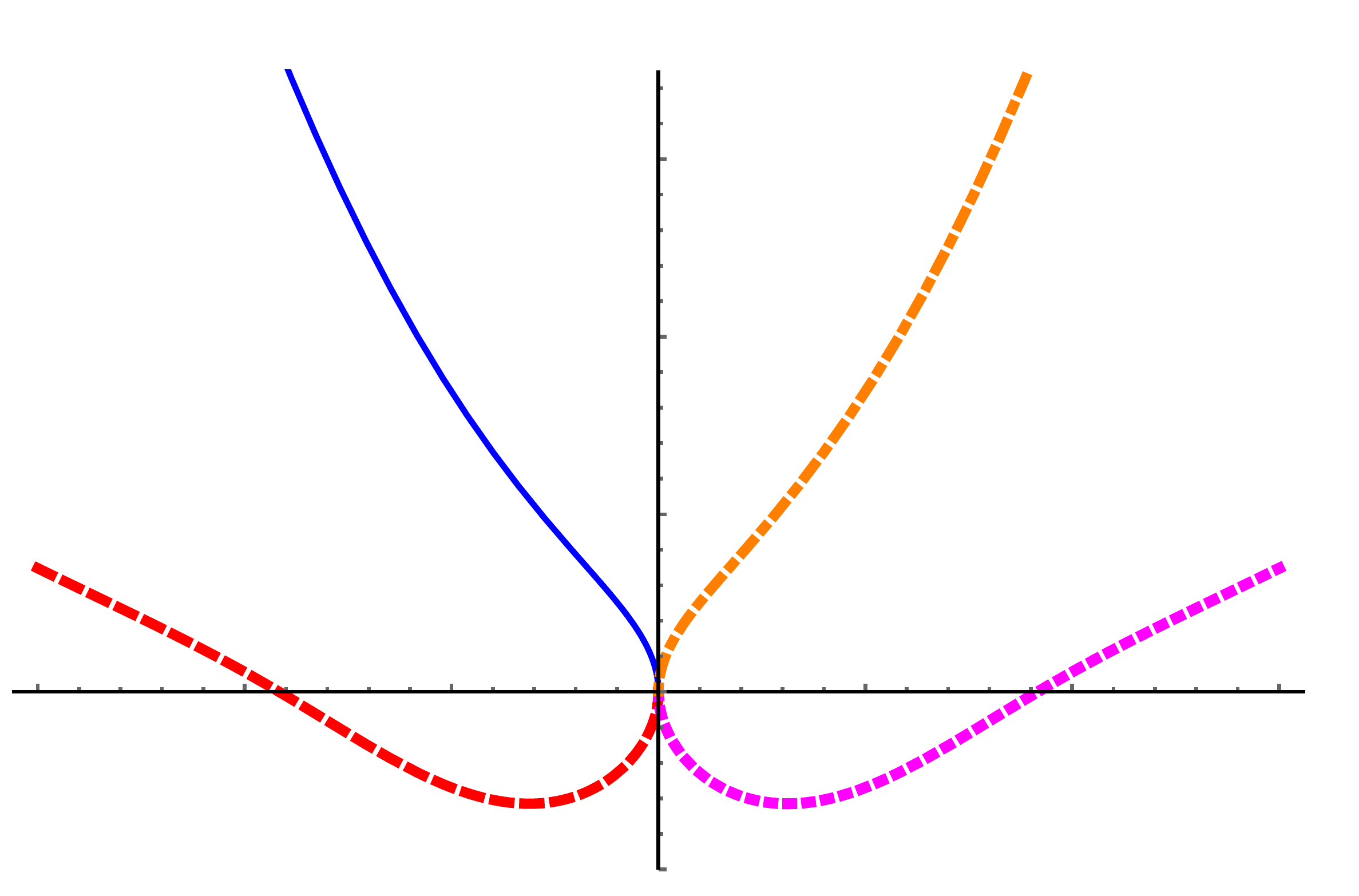}
\put(-150,170){\makebox(0,0)[lb]{\small$V_1(x)$}}
\put(-149,146){\makebox(0,0)[lb]{\small$3$}}
\put(-149,109){\makebox(0,0)[lb]{\small$2$}}
\put(-148.3,73){\makebox(0,0)[lb]{\small$1$}}
\put(-156,3){\makebox(0,0)[lb]{\small$-1$}}
\put(-7,39){\makebox(0,0)[lb]{\small$x$}}
\put(-282,30){\makebox(0,0)[lb]{\small$-1.5$}}
\put(-239,30){\makebox(0,0)[lb]{\small$-1.0$}}
\put(-197,30){\makebox(0,0)[lb]{\small$-0.5$}}
\put(-107,31){\makebox(0,0)[lb]{\small$0.5$}}
\put(-65,31){\makebox(0,0)[lb]{\small$1.0$}}
\put(-23,31){\makebox(0,0)[lb]{\small$1.5$}}
\caption{Case $N=5$: the $x$-component $V_1(x)$ of the doubly exotic potential $V(x,y)=V_1(x)+V_2(y)$ of type $Q_1$. It corresponds to the fifth-order integral ${\mathcal Y}_5$ (\ref{Y5Q1}). From the algebraic equations (\ref{AEQ1}) we obtain the four solutions $V_{1,i}(x) =T'_{1,i}$, $i=1,2,3,4$, displayed above. In the case $Q_1$ the $y$-component~$V_2(y)$ is of the same form with four similar solutions $V_{2,i}(y)$. The values $\beta_4=\delta=1$ were used. }
\end{figure}

\begin{figure}[h]\centering
\vspace{2ex}
\includegraphics[scale=0.4]{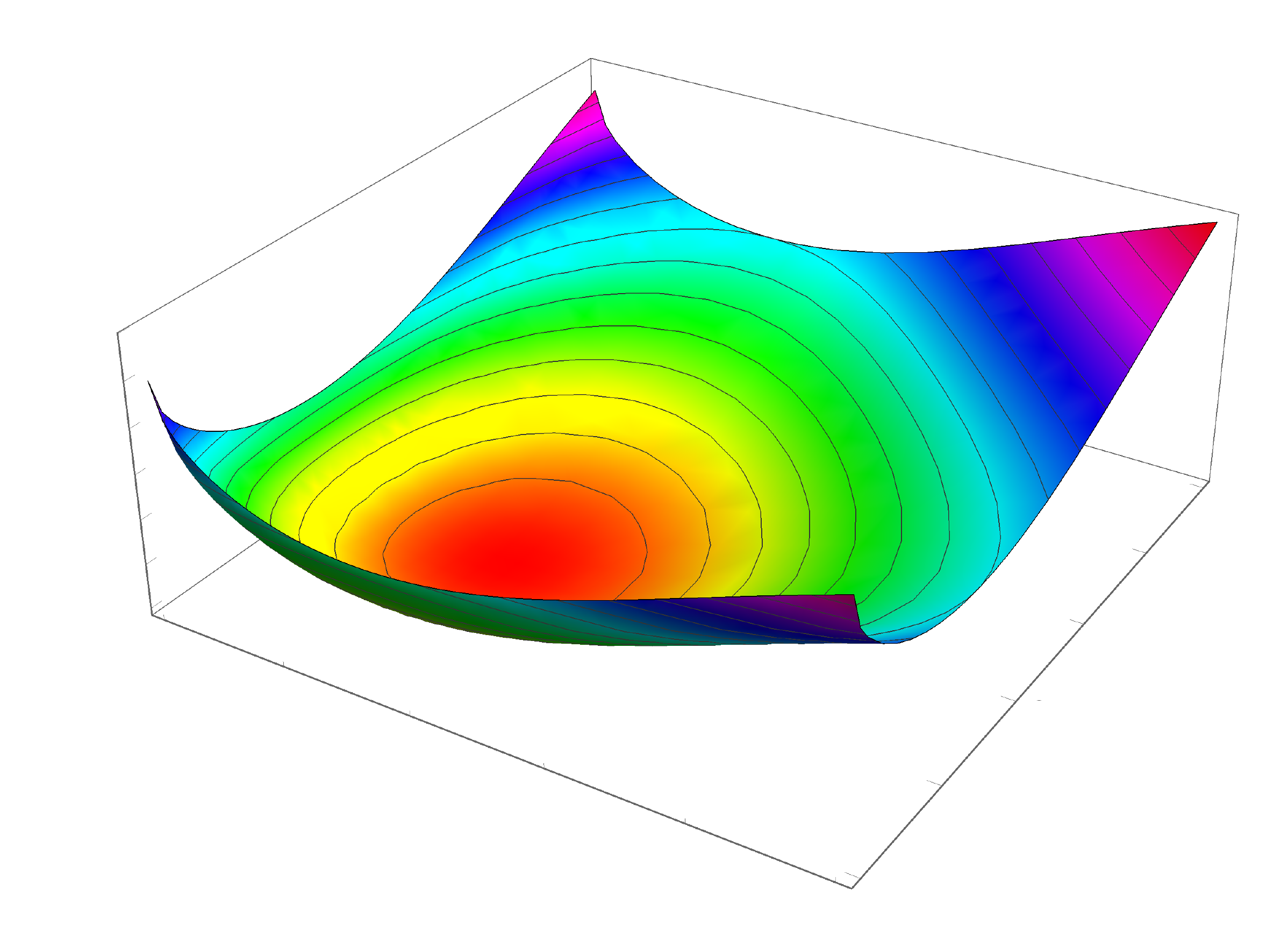}
\put(-170,186){\makebox(0,0)[lb]{\small$V=V_1(x)+V_2(y)$}}
\put(-264,58){\makebox(0,0)[lb]{\small$-1.25$}}
\put(-260,68,5){\makebox(0,0)[lb]{\small$-0.1$}}
\put(-267,77,5){\makebox(0,0)[lb]{\small$-0.75$}}
\put(-263,87,5){\makebox(0,0)[lb]{\small$-0.5$}}
\put(-269,97){\makebox(0,0)[lb]{\small$-0.25$}}
\put(-251,108){\makebox(0,0)[lb]{\small$0$}}
\put(-265,117){\makebox(0,0)[lb]{\small$0.25$}}
\put(-236,51){\makebox(0,0)[lb]{\small$0$}}
\put(-215,40){\makebox(0,0)[lb]{\small$0.2$}}
\put(-188,29){\makebox(0,0)[lb]{\small$0.4$}}
\put(-200,20){\makebox(0,0)[lb]{\small$x$}}
\put(-159,18){\makebox(0,0)[lb]{\small$0.6$}}
\put(-128,6){\makebox(0,0)[lb]{\small$0.8$}}
\put(-96,-6){\makebox(0,0)[lb]{\small$1.0$}}
\put(-80,0){\makebox(0,0)[lb]{\small$0$}}
\put(-62,20){\makebox(0,0)[lb]{\small$0.2$}}
\put(-47,38){\makebox(0,0)[lb]{\small$0.4$}}
\put(-17,40){\makebox(0,0)[lb]{\small$y$}}
\put(-31,56){\makebox(0,0)[lb]{\small$0.6$}}
\put(-18,71){\makebox(0,0)[lb]{\small$0.8$}}
\put(-7,84){\makebox(0,0)[lb]{\small$1.0$}}
\caption{A doubly exotic potential $Q_1$ corresponding to $N=5$. It admits the fifth-order integ\-ral~${\mathcal Y}_5$~(\ref{Y5Q1}). It also possesses bounded trajectories which by construction are closed and periodic. The values $\beta_4=\delta=1$ were used.}
\end{figure}

The algebra generated by the integrals takes the form
\begin{gather*}
\{C,{\mathcal X}\}_{\rm PB} = - 24\beta_4{\mathcal Y}_5^{(Q_1)},
\\
\big\{C,{\mathcal Y}_5^{(Q_1)} \big\}_{\rm PB} = 12X\big({\mathcal H}^2-{\mathcal X}^2\big)^2 - 48\delta{\mathcal X}{\mathcal H}.
\end{gather*}
In the corresponding quantum system analyzed in~\cite{AW}, the case $(i)$ splits into two subclasses of integrals ${\mathcal Y}_5$ that solely differ in their lower order $\hbar$-dependable terms. Consequently, two systems called $Q_1$ and $Q_2$ occur. However, in the classical limit $\hbar \rightarrow 0$ the two systems $Q_1$ and~$Q_2$ coincide.

Next, within case $(ii)$ the classical systems $Q_3$ ($A_{023}A_{050}A_{005}\neq 0$, $A_{122}=A_{032}=0$) and $Q_4$ ($A_{023}A_{005}\neq 0$, $A_{122}=A_{050}=A_{032}=0$) are \emph{not} superintegrable (like in the quantum case).

\medskip

Case $(ii)$.

$\bullet$ The system $Q_5$: $A_{023}\neq 0$, $A_{122}=A_{032}=A_{005}=0$, $A_{050}$ arbitrary.
This system corresponds to $A_{023}=1$ and arbitrary $A_{050}$, all other $A_{ijk}=0$. Again, by solving all the DE (\ref{D1})--(\ref{D3}) we arrive to the first-order nonlinear ODE for $T_1$
\begin{gather}\label{VQ5}
5 A_{050} (T_1')^3 - 12 \tau^2 x T_1' + 3\beta _1 (T_1')^2 - 12 \tau^2 T_1 + \mu = 0,
\end{gather}
$\tau \neq 0$, $\beta_1$ and $\mu$ are real constants, whereas
\begin{gather*}
 V_2 \equiv T_2' = \pm2\tau\sqrt{-y}.
\end{gather*}
The corresponding highest-order integral of motion reads
\begin{gather}
{\mathcal Y}_5^{(Q_5)} = A_{050}p_1^5 + p_1^2p_2^3 + p_1^3 (5 A_{050} T_1'+\beta_1) + p_1 \bigg(\frac{15}{2} A_{050} (T_1'){}^2+3\beta_1 T_1'-6 x \tau^2\bigg)\nonumber
\\ \hphantom{{\mathcal Y}_5^{(Q_5)} =}
{} + 6\tau\sqrt{-y} p_2 p_1^2 + 2T_1' p_2^3 + 12\tau\sqrt{-y} T_1' p_2.
\label{Y5Q5}
\end{gather}
Clearly, the case $A_{032}=1$ and arbitrary $A_{005}$ (all other $A_{ijk}=0$) also leads to a superintegrable potential. It can simply be obtained by replacing $A_{050} \rightarrow A_{005}$ and making the permutation $x \Leftrightarrow y$ ($V_1 \Leftrightarrow V_2$) in (\ref{VQ5}) and (\ref{Y5Q5}).

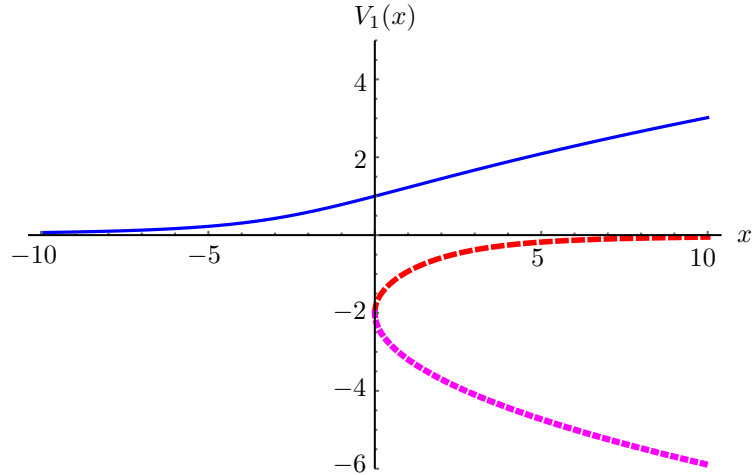
\begin{figure}[h!]\centering
\includegraphics[scale=0.4]{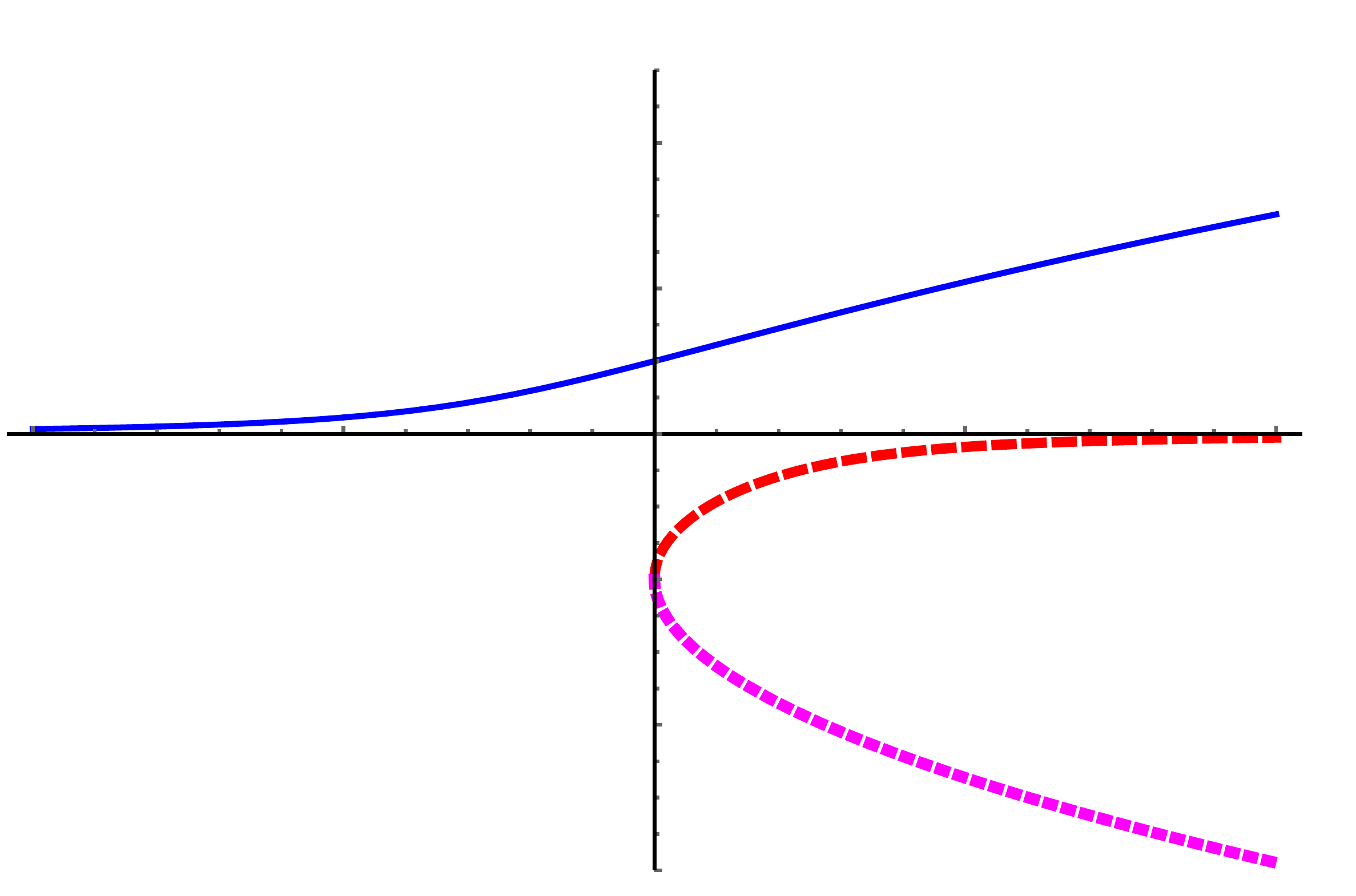}
\put(-150,171){\makebox(0,0)[lb]{\small$V_1(x)$}}
\put(-150,148){\makebox(0,0)[lb]{\small$4$}}
\put(-150,119){\makebox(0,0)[lb]{\small$2$}}
\put(-158,61){\makebox(0,0)[lb]{\small$-2$}}
\put(-158,32){\makebox(0,0)[lb]{\small$-4$}}
\put(-158,3){\makebox(0,0)[lb]{\small$-6$}}
\put(-5,91){\makebox(0,0)[lb]{\small$x$}}
\put(-280,82){\makebox(0,0)[lb]{\small$-10$}}
\put(-213,82){\makebox(0,0)[lb]{\small$-5$}}
\put(-83,82){\makebox(0,0)[lb]{\small$5$}}
\put(-23,82){\makebox(0,0)[lb]{\small$10$}}
\caption{Case $N=5$: the $x$-component $V_1(x)$ of the doubly exotic potential $V(x,y)=V_1(x)+V_2(y)$ of type $Q_5$. It admits the fifth-order integral ${\mathcal Y}_5$ (\ref{Y5Q5}). From the equation (\ref{VQ5}) we obtain the three numerical solutions $V_{1,i}(x) =T'_{1,i}$, $i=1,2,3$, displayed above. The values $A_{050}=\frac{1}{5}$, $\tau=\frac{1}{\sqrt{12}}$, $\beta_1=1$ and $\mu=-3$ were used.}
\end{figure}

{\samepage
In this case, the algebra generated by the integrals $C=\big\{{\mathcal Y}_5^{(Q_5)},{\mathcal X}\big\}_{\rm PB}$, ${\mathcal Y}_5^{(Q_5)}$ and $\mathcal X$ takes the form
\begin{gather*}
\{C,{\mathcal X}\}_{\rm PB} = 0,
\\
\big\{ C,{\mathcal Y}_5^{(Q_5)} \big\}_{\rm PB} = -144\tau^4{({\mathcal H}+{\mathcal X})},
\end{gather*}
and it does not provide further information about the solutions of (\ref{VQ5}).

}

$\bullet$ The system $Q_6$: $A_{023}A_{032}\neq 0$, $A_{122}=A_{050}=A_{005}=0$.
This system corresponds to $A_{023} A_{032}\neq 0$, all other $A_{ijk}=0$.
\begin{gather}
3 A_{032} (T_1'){}^2 - \sigma _2 x T_1' - \sigma _2 T_1 = 0,\nonumber
\\
3 A_{023} (T_2'){}^2 + \sigma _2 y T_2' + \sigma _2 T_2 = 0,
\label{VQ6}
\end{gather}
$\sigma _2 \neq 0$ is a real constant.
\begin{gather}
{\mathcal Y}_5^{(Q_6)} = A_{032}p_2^2 p_1^3 + A_{023}p_2^3 p_1^2 + 2 p_2^3 A_{023} T_1' + 2 p_1^3 A_{032} T_2' + p_1 p_2^2 \bigg(3 A_{032} T_1'-\frac{\sigma _2 x}{2}\bigg)\nonumber
\\ \hphantom{{\mathcal Y}_5^{(Q_6)} =}
{}+ p_1^2 p_2 \bigg(3 A_{023} T_2'+\frac{\sigma _2 y}{2}\bigg) + p_1 \bigg(3 x A_{023} T_2' T_2''+6 A_{032} T_1' T_2'+\frac{1}{2} \sigma _2 x y T_2''\bigg)\nonumber
\\ \hphantom{{\mathcal Y}_5^{(Q_6)} =}
{}+ p_2 \bigg(3 y A_{032} T_1' T_1''+6 A_{023} T_1' T_2'-\frac{1}{2} \sigma _2 x y T_1''\bigg).
\label{Y5Q6}
\end{gather}

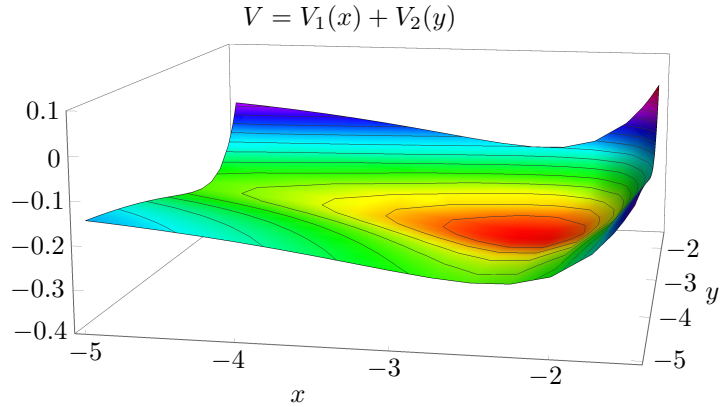
\begin{figure}[h]\centering
\vspace{2ex}
\includegraphics[scale=0.4]{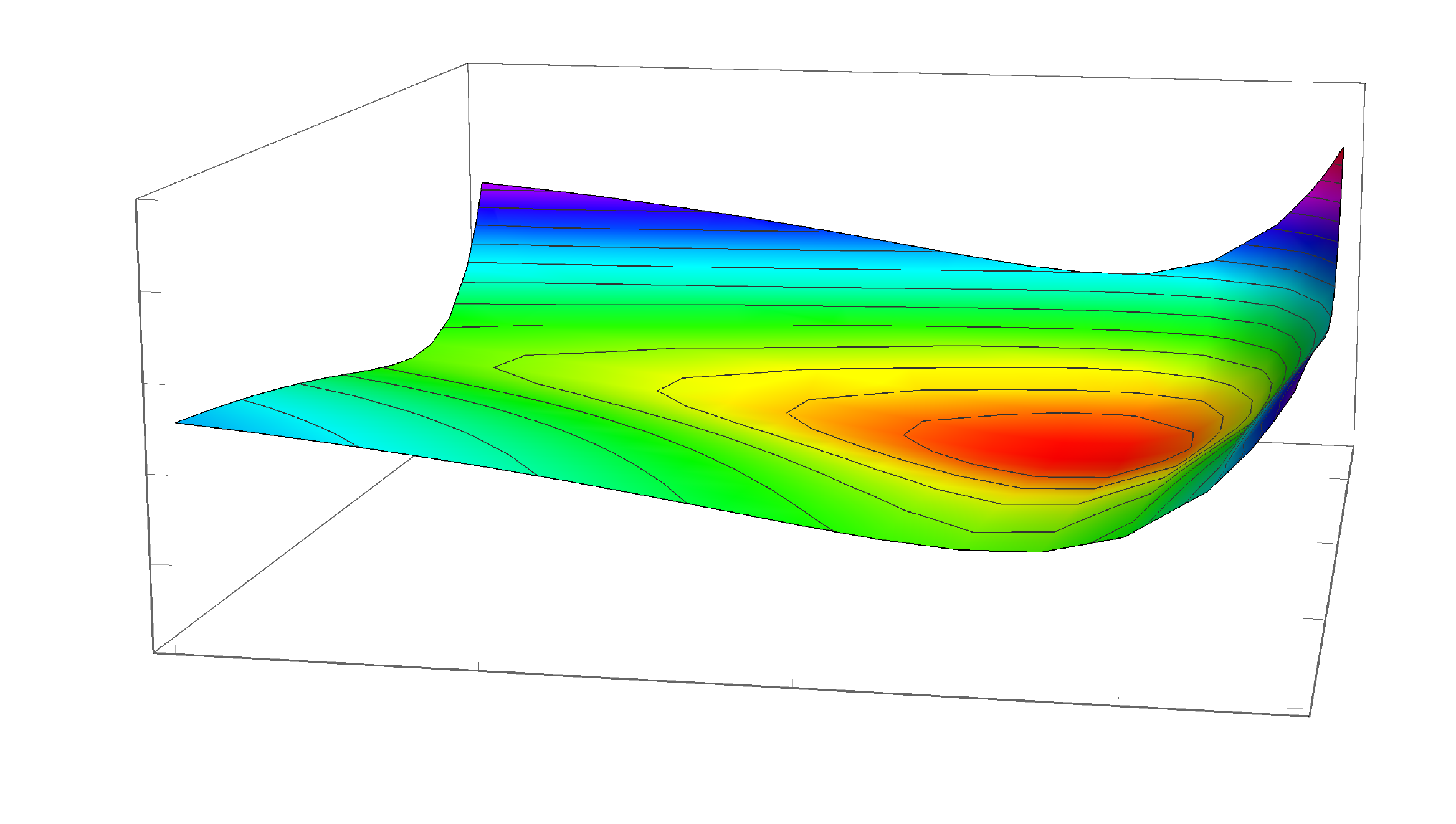}
\put(-165,131){\makebox(0,0)[lb]{\small$V=V_1(x)+V_2(y)$}}
\put(-252,14){\makebox(0,0)[lb]{\small$-0.4$}}
\put(-253,30){\makebox(0,0)[lb]{\small$-0.3$}}
\put(-253,46){\makebox(0,0)[lb]{\small$-0.2$}}
\put(-253,63){\makebox(0,0)[lb]{\small$-0.1$}}
\put(-239,80){\makebox(0,0)[lb]{\small$0$}}
\put(-247,98){\makebox(0,0)[lb]{\small$0.1$}}
\put(-231,6){\makebox(0,0)[lb]{\small$-5$}}
\put(-177,3){\makebox(0,0)[lb]{\small$-4$}}
\put(-147,-9){\makebox(0,0)[lb]{\small$x$}}
\put(-118,0){\makebox(0,0)[lb]{\small$-3$}}
\put(-57,-3){\makebox(0,0)[lb]{\small$-2$}}
\put(-10,3){\makebox(0,0)[lb]{\small$-5$}}
\put(-8,19){\makebox(0,0)[lb]{\small$-4$}}
\put(-6,32){\makebox(0,0)[lb]{\small$-3$}}
\put(-5,45){\makebox(0,0)[lb]{\small$-2$}}
\put(10,28){\makebox(0,0)[lb]{\small$y$}}
\caption{A doubly exotic potential $Q_6$ corresponding to $N=5$. It admits the fifth-order integral~${\mathcal Y}_5$~(\ref{Y5Q6}). It also possesses bounded trajectories which by construction are closed and periodic. The values $A_{032}=-A_{023}=\frac{1}{12}$, $\sigma_1=1$ were used.}
\end{figure}

In this case, the algebra generated by the integrals $C=\big\{{\mathcal Y}_5^{(Q_6)},{\mathcal X}\big\}_{\rm PB}$, ${\mathcal Y}_5^{(Q_6)}$ and $\mathcal X$ takes the form
\begin{gather*}
\{ C,{\mathcal X}\}_{\rm PB} = 0,
\\
\big\{C,{\mathcal Y}_5^{(Q_6)} \big\}_{\rm PB} = 2{ \mathcal X}\sigma_2^2\big({\mathcal H}^2 - {\mathcal X}^2\big)^2,
\end{gather*}
and it does not provide further information about the solutions of (\ref{VQ6}). However, it is easy to check that
\begin{gather*}
 T_1(x) = \frac{W^2(x) - \sigma_2 x^2}{12 A_{032}},
\qquad
 T_2(y) = -\frac{W^2(y) - \sigma_2 y^2}{12 A_{023}},
\end{gather*}
satisfy (\ref{VQ6}), where $W=W(z)$ is given by the following third order polynomial equation
\begin{gather*}
 \big(W - z \sqrt{\sigma_2}\big)\big(W + 2 z \sqrt{\sigma_2}\big)^2 + \tau = 0,
\end{gather*}
here $\tau \neq 0$ is an integration constant.

$\bullet$ The system $Q_7$.
This system is a particular case of system $Q_5$. It corresponds to the situation where $A_{023}=1$ and all other $A_{ijk}=0$.
\begin{gather*}%\label{VQ7}
 3\beta _1 (T_1'){}^2 - 12 \tau^2 x T_1' - 12 \tau^2 T_1 + \mu = 0,\qquad
V_2 \equiv T_2' = \pm2 \tau \sqrt{-y},
\end{gather*}
$\tau \neq 0$, $\beta_1$ and $\mu$ are real constants.
\begin{gather*}
{\mathcal Y}_5^{(Q_7)} = p_1^2 p_2^3 + p_1^3\beta_1 +
p_1 \big(3\beta_1 T_1'-6 x \tau^2\big)
 + 6\tau\sqrt{-y} p_2 p_1^2+ 2 T_1'p_2^3 + 12\tau\sqrt{-y} T_1'p_2.
\end{gather*}

\pagebreak

{\it Case} $(iii)$.

$\bullet$ The system $Q_8$: $A_{050}^2+A^2_{005}\neq 0$, $A_{122}=A_{023}=A_{032}=0$.
This system corresponds to the case $A_{050}^2+A^2_{005}\neq 0$ and all other $A_{ijk}=0$.
\begin{gather}
A_{050} T_1'{}^3 + \beta _1 T_1'{}^2 + \theta _1 T_1' - \Lambda x + \kappa _1 = 0,\nonumber
\\
A_{005} T_2'{}^3 + \alpha _1 T_2'{}^2 + \phi_1 T_2' + \Lambda y + \omega _1 = 0,
\label{VQ8}
\end{gather}

$\Lambda \neq 0$, $\alpha _1,\beta_1,\theta _1,\kappa _1,\phi_1$ and $\omega _1$ are real constants.
\begin{gather*}
{\mathcal Y}_5^{(Q_8)} = A_{050}p_1^5 + A_{005}p_2^5 + p_1^3 \bigg(5 A_{050} T_1'+\frac{5 \beta _1}{3}\bigg) + p_2^3 \bigg(5 A_{005} T_2'+\frac{5 \alpha _1}{3}\bigg)
\\ \hphantom{{\mathcal Y}_5^{(Q_8)} = }
{}+ p_1 \bigg(\frac{15}{2} A_{050} T_1'{}^2+5 \beta _1 T_1'+\frac{5 \theta _1}{2}\bigg) + p_2 \bigg(\frac{15}{2} A_{005} T_2'{}^2+5 \alpha _1 T_2'+\frac{5 \phi _1}{2}\bigg).
\end{gather*}

In this case, for the solutions of (\ref{VQ8}) the Poisson bracket $C = \big\{{\mathcal Y}_5^{(Q_8)},{ \mathcal X} \big\}_{\rm PB} \propto \lambda$. Thus, the algebra of the integrals of motion takes the form
\begin{equation*}
\{C,{ \mathcal X}\}_{\rm PB} = 0, \qquad
\big\{ C,{\mathcal Y}_5^{(Q_8)}\big\}_{\rm PB} = 0.
\end{equation*}
Again, the system $Q_8$ was found in~\cite{GungorKNN:2014} by means of Heisenberg-type higher order symmetries. All three conserved quantities $\big({\mathcal H},{\mathcal X},{\mathcal Y}_5^{(Q_8)}\big)$ admit separation of variables in Cartesian coordinates. The particular case with $\beta_1=\theta_1=\kappa_1=\alpha_1=\phi_1
=\omega_2=0$, thus $V_1 \propto x^{\frac{1}{3}}$ and $V_2 \propto y^{\frac{1}{3}}$, was studied in~\cite{Grigoriev} using action-angle variables.

$\bullet$ The system $Q_9$.
This system is a particular case of system $Q_8$. It corresponds to the situation where $A_{050}=1$ and all other $A_{ijk}=0$.
\begin{gather*}
 A_{050} T_1'{}^3 + \beta _1 T_1'{}^2 + \theta _1 T_1' - \Lambda x + \kappa _1 = 0,\nonumber
\\
\alpha _1 T_2'{}^2 + \phi_1 T_2' + \Lambda y + \omega _1 = 0,
\label{VQ9}
\end{gather*}

$\Lambda \neq 0$, $\beta_1$, $\theta_1$, $\kappa_1$, $\phi_1$, $\omega _1$ and $\alpha_1^2+\phi_1^2 \neq0$ are real constants.
\begin{gather*}
{\mathcal Y}_5^{(Q_9)} = A_{050}p_1^5 + p_1^3 \bigg(5 A_{050} T_1'+\frac{5 \beta _1}{3}\bigg) + \frac{5 \alpha _1}{3}p_2^3
+ p_1 \bigg(\frac{15}{2} A_{050} T_1'{}^2+5 \beta _1 T_1'+\frac{5 \theta _1}{2}\bigg)
\\ \hphantom{{\mathcal Y}_5^{(Q_9)} =}
{}+ p_2 \bigg(5 \alpha _1 T_2'+\frac{5 \phi _1}{2}\bigg).
\end{gather*}

\section{Conclusions}\label{conclu}

We considered $N$th-order superintegrable classical systems in a two-dimensional Euclidean space separating in Cartesian coordinates. They are characterized by three polynomial (in momentum variables) integrals of motion (${\mathcal H},{\mathcal X},{\mathcal Y}_N$). Let us summarize the main results reported in this paper:
\begin{enumerate}\itemsep=0pt
\item Higher-order ($N>2$) superintegrable classical systems
\[
{\mathcal H} = \frac{1}{2}\big(p_1^2 + p_2^2\big) + V_1(x) + V_2(y),
\]
can be classified into three classes: doubly standard, singly exotic and doubly exotic potentials. This classification is based on the nature of the equation that defines the most general form of the potential functions $V_1(x)$ and $V_2(x)$. For doubly standard potentials this equation is a linear compatibility condition necessary for the existence of the $N$th order integral of motion ${\mathcal Y}_N$ (in general a PDE of order $N$ in two variables) whilst in the case of doubly exotic potentials it is given by a nonlinear compatibility condition.

\item From the equation $\{{\mathcal Y}_N,{\mathcal H}\}_{\rm PB}=0$, we show in a systematic manner how to find and successively solve a “well” of NLCC separately for $V_1(x)$ and $V_2(y)$. It was also indicated that requiring the integrals of motion (${C}=\{{\mathcal X},{\mathcal Y}_N\}_{\rm PB},{\mathcal X},{\mathcal Y}_N$) to span a closed polynomial algebra may help to simplify (reduce the order) of the DE and eventually to find the explicit solutions $V_1(x)$ and $V_2(y)$.

\item All fifth-order ($N=5$) superintegrable doubly exotic potentials were derived explicitly by solving the set of DE. {The DE lead to first order non-linear ODEs that define the functions $V_1$ and $V_2$, respectively. Unlike the quantum case, these equations do not have the Painlev\'e property. This was verified either by finding their general solutions explicitly or by applying a standard test to them~\cite{Baldwin}}. Interestingly, at $N=5$ doubly exotic confining potentials appear for the first time. At $N=4$ no doubly exotic potentials occur at all.

\item The present study suggests to explore the inverse problem, namely we take two polynomial functions $({\mathcal A}$ and ${\mathcal B})$ in momentum variables ($p_1,p_2$) and construct the new object ${\mathcal C}=\{{\mathcal A},{\mathcal B}\}_{\rm PB}$. If the algebra generated by $({\mathcal A},{\mathcal B},{\mathcal C})$ is a closed polynomial algebra with polynomial coefficients in ${\mathcal H}
$, then \textit{under what conditions this closure relations imply that~${\mathcal A}$ and~${\mathcal B}$ are integrals, i.e., they Poisson commute with~$\mathcal H$}?
\end{enumerate}

Finally, a direct computation for the next two cases $N=6,7$ leads us to the following conjecture:

\begin{Conjecture*}
There exists an infinite family of $N$th-order superintegrable systems with an inte\-gral
\begin{gather*}
{\mathcal Y}_{N}^{(\rm Doubly\ exotic)} = L^{(N-4)}_zp_1^{2}p_2^{2} + {\rm(lower\ order\ terms)}, \qquad N\geq 5,
\\
\big\{{\mathcal Y}_{N}^{(\rm Doubly\ exotic)},{\mathcal H}\big\}_{\rm PB} = 0.
\end{gather*}
The associated potential $V$ can be written as follows
\begin{gather*}
V = V_1(x) + V_2(y) = \mathcal{G}'(x;N) + \mathcal{G}'(y;N),
\end{gather*}
here $\mathcal{G}=\mathcal{G}(u;N)$ obeys a nonlinear first-order ODE of the form
\begin{gather}
{\mathcal{G}}'\big[6u^{N-4}{\mathcal{G}}' +4{(N-5)}u^{N-5}{\mathcal{G}} + F_{1}(u) + \sigma u^{N-2} \big] \nonumber
\\ \qquad
 {}+ {\mathcal{G}}\big[ 2{(N-5)}u^{N-6} {\mathcal{G}} +F_{2}(u) -2\sigma u^{N-3} \big] + F_{3}(u) + bu^N = 0.
\label{F2N}
\end{gather}
The three functions $F_{q}$'s in \eqref{F2N} are polynomials in the variable $u$ of degree at most $(N-1)$, and $\sigma$, $b$ are real parameters as well. The equation~\eqref{F2N} is in complete agreement with the limit $\hbar \rightarrow 0$ of its quantum analogue treated in~{\rm \cite{AMGen}}. In future work, we plan to establish in detail under what conditions the closed algebra of the integrals of motion is polynomial, and how to use it as a new systematic tool to solve the determining equations in a simpler and more efficient manner.
\end{Conjecture*}

\subsection*{Acknowledgments}

\.{I}Y and AMER during a sabbatical leave and a postdoctoral academic stay at the Centre de Recherches
Math\'ematiques, Universit\'e de Montr\'eal, respectively, were introduced to the subject of higher-order superintegrability by Pavel Winternitz. His enormous influence is present in this study as it does in the whole subject. It is with admiration and great affection that we dedicate this paper to his memory. We~thank the anonymous referees and the editor for their valuable comments and constructive suggestions on the manuscript.

\pdfbookmark[1]{References}{ref}
\LastPageEnding

\end{document}